\input phyzzx.tex
\tolerance=1000
\voffset=-0.0cm
\hoffset=0.7cm
\sequentialequations
\def\rl{\rightline}

\def\t1{{\tilde 1}}

\def\t{\theta}

\REF{\HOL}{G. 't Hooft, [arXiv:gr-qc/9310026].}
\REF{\LEN}{L. Susskind, J. Math. Phys. {\bf 36} (1995) 6377, [arXiv:hep-th/9409089].} 
\REF{\RAP}{R. Bousso, Rev. Mod. Phys. {\bf 74} (2002) 825, [arXiv:hep-th/0203101].}
\REF{\BEKK}{J. D. Bekenstein, Lett. Nuov. Cimento {\bf 4} (1972) 737; Phys Rev. {\bf D7} (1973) 2333; Phys. Rev. {\bf D9} (1974) 3292.}
\REF{\HAW}{S. Hawking, Nature {\bf 248} (1974) 30; Comm. Math. Phys. {\bf 43} (1975) 199.}
\REF{\MAL}{J. Maldacena, Adv. Theor. Math. Phys. {\bf 2} (1998) 231, [arXiv:hep-th/9711200].} 
\REF{\GKP}{S. Gubser, I. Klebanov and A. Polyakov, Phys. Lett. {\bf B428} (1998) 105, [arXiv:hep-th/9802109].} 
\REF{\WIT}{E. Witten, Adv. Theor. Math. Phys. {\bf 2} (1998) 253, [arXiv:hep-th/9802150].}
\REF{\VER}{E. P. Verlinde, arXiv:1001.0785[hep-th].}
\REF{\CUL}{H. Culetu, arXiv:1002.3876[gr-qc]; arXiv:1005.1570[gr-qc].}
\REF{\LEE}{J. W. Lee, arXiv:1003.4464[hep-th]; arXiv:1011.1657[hep-th].}
\REF{\WAL}{R. M. Wald, "General Relativity", The University of Chcago Press, 1984.}
\REF{\SUS}{L. Susskind, arXiv:1101.6048[hep-th].}
\REF{\FUR}{D. V. Fursaev, Phys. Rev. {\bf D77} (2008) 124002, arXiv:0711.1221[hep-th]; Phys.Rev. {\bf D82} (2010) 064013, arXiv:1006.2623[hep-th].}
\REF{\UNR}{W. G. Unruh, Phys. Rev. {\bf D14} (1976) 870.}
\REF{\PAD}{T. Padmanabhan, Class. Quant. Grav. {\bf 21} (2004) 4485, [arXiv:gr-qc/0308070].}
\REF{\JAC}{T. Jacobson, Phys. Rev. Lett. {\bf 75} (1995) 1260, [arXiv:gr-qc/9504004].}
\REF{\JR}{T. Jacobson and R. Parenti, Found. Phys. {\bf 33 (2)} (2003) 323, [arXiv:gr-qc/0302099].}
\REF{\BEK}{J. D. Bekenstein, Phys. Rev. {\bf D23} (1981) 287.}
\REF{\BOU}{R. Bousso, JHEP {\bf 0405} (2004) 050, [arXiv:hep-th/0402058.}
\REF{\GCEB}{E. E. Flanagan, D. Marolf and R. M. Wald, Phys. Rev. {\bf D62} (2000) 084035, [arXiv: hep-th/9908070].}
\REF{\LAST}{E. Halyo, JHEP {\bf 1004} (2010) 097,  arXiv:0906.2164[hep-th].}
\REF{\MIL}{M. Milgrom, Astrophys. Journ. {\bf 270} (1983) 365; ibid {\bf 270} (1983) 371; ibid {\bf 270} (1983) 384.} 
\REF{\BEKE}{J. D. Bekenstein, Phys. Rev. {\bf D70 (8)} (2004) 083509,  [arXiv:astro-ph/0403694].}
\REF{\LENN}{L. Susskind, [arXiv:hep-th/9309145].}
\REF{\SBH}{E. Halyo, A. Rajaraman and L. Susskind, Phys. Lett. {\bf B392} (1997) 319, [arXiv:hep-th/9605112].}
\REF{\HRS}{E. Halyo, B. Kol, A. Rajaraman and L. Susskind, Phys. Lett. {\bf B401} (1997) 15, [arXiv:hep-th/9609075].}
\REF{\EDI}{E. Halyo, Int. Journ. Mod. Phys. {\bf A14} (1999) 3846, [arXiv:hep-th/9610068]; Mod. Phys. Lett. {\bf A13} (1998) 1521,
[arXiv:hep-th/9611175]; [arXiv:hep-th/0107169]; JHEP {\bf 0112} (2001) 005, [arXiv:hep-th/0108167].}
\REF{\PADD}{T. Padmanabhan, Phys. Rept. {\bf 406} (2005) 49, [arXiv:gr-qc/0311036]; Rept. Prog. Phys. {\bf 73} (2010) 046901, 
arxiv:0911.5004[gr-qc].}

\singlespace
\rl{SU-ITP-11/19}
\pagenumber=0
\normalspace
\medskip
\bigskip
\titlestyle{\bf{Entropic Gravity in Rindler Space}}
\smallskip
\author{ Edi Halyo{\footnote*{e--mail address: halyo@stanford.edu}}}
\smallskip
\centerline {Department of Physics} 
\centerline{Stanford University} 
\centerline {Stanford, CA 94305}
\smallskip
\vskip 2 cm
\titlestyle{\bf ABSTRACT}
We show that Rindler horizons are entropic screens and gravity is an entropic force in Rindler space by deriving the Verlinde entropy formula from the focusing of light due to a mass close to the horizon. Consequently, gravity is also entropic in the near horizon regions of Schwarzschild and de Sitter space-times. In different limits, the entropic nature of gravity in Rindler space leads to the Bekenstein entropy bound and the uncertainty principle.

\singlespace
\vskip 0.5cm
\endpage
\normalspace

\centerline{\bf 1. Introduction}
\medskip

Of the four fundamental forces of Nature, gravity though universal, is the least understood and the most mysterious one. This is mainly due to the fact that, in spite of all the efforts, gravity as described by General Relativity has not been quantized yet. In fact, for this reason and others, it is clear that gravitation is different from all other interactions.
String theory notwithstanding, it seems that we are missing some of the fundamental properties of gravity that are essential for a deeper understanding that will hopefully lead to its quantization. For example, nowadays it is widely believed that holography[\HOL,\LEN,\RAP] which was inspired by the Bekenstein-Hawking entropy of black holes
[\BEKK,\HAW] and manifestly realized by the AdS/CFT duality[\MAL,\GKP,\WIT] is such a property.
The discovery of new properties of gravity is extremely important in our quest to understand it at a
more fundamental level.

In ref. [\VER] Verlinde argued that gravity is an entropic force that by definition satisfies 
$$F=T {{\Delta S} \over {\Delta x}}\eqno(1)$$
i.e. the entropic force is proportional to the local temperature and in the direction of the gradient of the screen entropy. From this
viewpoint, a mass gravitationally attracts another because its motion towards a hypothetical screen (close to it) increases the screen entropy. In addition
to saturating the holographic entropy bound, these screens satisfy a formula that describes how their entropy changes as a function
of the distance, $\Delta x$, at which a mass is located
$$\Delta S={{2 \pi kmc} \over \hbar} \Delta x \eqno(2)$$
The entropic nature of force leads to Newton's second law since the temperature and the entropy gradient are proportional to the acceleration and mass respectively.
In ref. [\VER], eq. (2) was not derived but basically postulated based on some reasonable arguments. In addition,
the postulated screens which attract masses were not identified except for their two properties mentioned above.

In this paper, we argue that these screens are Rindler horizons and gravity (both Newtonian and General Relativity) behaves as an entropic force in Rindler space.{\footnote1{This was previously claimed in refs. [\CUL,\LEE] but the arguments in those works seem
to be incorrect even though the correct formulas were derived. See the discussion at the end of section 2.}
First, if a mass
is accelerating due to a force acting on it, we can always go to its (accelerating) rest frame in which there is a Rindler horizon perpendicular to the direction of the acceleration. Thus, Rindler horizons are natural screens in accelerated frames.
Second, Rindler horizons saturate the holographic entropy bound by definition. Finally,
and maybe most importantly, we can derive the Verlinde entropy formula in eq. (2) from the focusing of light due to a mass above the horizon. The same result can be obtained (up to a factor of two) by considering the change in the shape of the horizon due to the
gravitational field of the mass. Since we use gravity, i.e. General Relativity, to derive the Verlinde entropy formula, this cannot be used in turn to derive the gravitational force. However, our results constitute a nontrivial consistency check, albeit in Rindler space, of the gravitational force, its entropic origin and the Verlinde entropy formula.

This immediately implies that gravity behaves as an entropic force in all space-times that reduce to Rindler space such as the near horizon regions of
Schwarzschild black holes, de Sitter spaces and all black branes. In these cases, the Rindler horizon coincides with the 
respective event horizons and the Unruh and Hawking temperatures match. Unfortunately, our results cannot be easily generalized to regions far from the horizons or simply to space-times without horizons.

We also show that the Verlinde entropy formula leads to the Bekenstein entropy bound for weakly gravitating
systems by deriving the latter as a limiting case of the former. In addition, we relate the focusing of light due to a mass to the uncertainty principle by considering the smallest entropy change possible on the horizon.
Finally, we entertain some speculative ideas such as upper and lower bounds on acceleration and
the holographic description of Rindler space.

The paper is organized as follows. In Section 2, we derive eq. (2) in Rindler space and use it and the horizon thermodynamics to obtain Newtonian gravity and General Relativity in Rindler space. In Section 3, we describe the corollaries of our results including gravity in near horizon geometries, a derivation of the Bekenstein entropy bound and the relation to the uncertainty principle. In Section 4, we consider some speculative ideas related to the entropic nature of gravity in Rindler space. Finally, Section 4 contains our conclusions and a discussion of our results.

\bigskip
\centerline{\bf 2. Gravity as an Entropic Force in Rindler Space}
\medskip

In order to understand the entropic origin of Newton's second law and gravity, we consider a mass $m$ in Minkowski space (with coordinates $T,X,Y,Z$) with an acceleration $\vec a$
due to a force $\vec F$. We can always go to the rest frame of the object which is an accelerated frame. The coordinate transformation
between Minkowski space and the accelerated frame is, assuming $\vec a=(a,0,0)$,
$$cT=x \sinh \left({{at} \over c} \right) \qquad X=x \cosh \left({{at} \over c} \right) \eqno(3)$$
and $Y=y$, $Z=z$. The coordinates $t,x,y,z$ describe Rindler space which is how an accelerated observer sees Minkowski space. 
A mass that is at rest in Rindler space corresponds to an accelerating mass in Minkowski space. There the mass starts at early times $T \sim -\infty$, to move from $X \sim \infty$ towards the origin, decelerating with $a$. At $T=0$ it reaches its closest distance from the origin at $X=c^2/a$ and momentarily stops. For $T>0$ it turns around and accelerates towards the positive $X$ direction, i.e. $X \to \infty$ as $T \to \infty$.

The Rindler space metric is 
$$ds^2=-(ax)^2 dt^2 +dx^2 +dy^2+dz^2 \eqno(4)$$
which is flat in the spatial directions. From this metric we see that there is a horizon, i.e. the Rindler horizon at $x=0$.
In these coordinates, the mass is at $x=c^2/a$ and $y=z=0$. 
In Minkowski space, the horizon is a light-sheet (defining the Rindler wedge in the $T$--$X$ plane) which simply arises due to the fact that light behind it cannot reach the accelerating object in finite time. 

We would like to show the entropic nature of Newton's second law and gravity in Rindler space. First of all, not only can we always
go to the rest frame of an accelerating object but, in this frame, there are natural candidates for the screens that Verlinde
postulated. These are the Rindler horizons which are infinite planes (in our case the $y$--$z$ plane) with a holographically saturated  
entropy density. The only property of screens that is not immediately manifest is the entropy change formula (for the screen) due to  masses close to them i.e. eq. (2). We derive this formula below by using the focusing of light due to the gravitational effect of the mass. We also give an alternative derivation using the gravitational effect of the mass on the area of the horizon.

Consider a mass of $m$ in Rindler space at $x=c^2/a$ or a distance of $\Delta x=c^2/a$ from the Rindler horizon. The mass is
at rest and therefore feels an acceleration $a$ towards the horizon. The Rindler horizon is the infinite
$y$--$z$ plane with maximum entropy density per transverse area. Therefore, the horizon entropy cannot increase by the increase of the area or entropy density. The best way to understand how the presence of the mass affects the horizon entropy is to consider a cylindrical sheet of light with radius (or impact parameter) $b$ shining towards the horizon. {\footnote2{I would like to thank Lenny Susskind for pointing out this definition of horizon entropy.}} 
The mass, due to its gravity, focuses the light sheet and therefore the radius of the light sheet is smaller when it hits the horizon. In Rindler coordinates, it takes light an infinite coordinate time but a finite affine parameter to reach the horizon. In any case, we can always consider the stretched horizon at $x \sim \ell_P$ which light reaches in finite coordinate time.

It is easier to calculate the focusing of light in Minkowski space since it is flat except for the gravitational effects of the mass. Clearly, in these coordinates light reaches the horizon in a finite time. 
We now consider a snapshot at $t=T=0$ when the mass is momentarily at rest and the coordinates $x$ and $X$ in the Rindler and Minkowski space-times coincide. At this moment,
the horizon and the mass are at $x=X=0$ and $\Delta x=X=c^2/a$ respectively. The metric
due to the mass is simply the Schwarzschild metric.
In General Relativity, the bending of light due to a mass is parametrized by the deflection angle
$\phi$. For weak gravity, i.e. for $Gm/b<<1$, the null geodesic equations in the Schwarzschild metric give the well-known result[\WAL] {\footnote3{Taking $m=E/c$ for light, Newtonian gravity gives half this result.}}
$$\phi={{4 Gm} \over {bc^2}} \eqno(5)$$
For small angles of deflection, $\phi \simeq \tan \phi$ and therefore the radius of the light sheet on the horizon is
$$b^{\prime}=b-{{4Gm} \over {bc^2}} \Delta x \eqno(6)$$
The area deficit on the Rindler horizon due to the mass is given by 
$$\Delta A=A^{\prime}-A \simeq -{{8 \pi Gm} \over c^2} \Delta x \eqno(7)$$
where, in addition to assuming weak gravity, we neglected the higher order term in $\Delta x/ b$. Note that the area decreases for 
$\Delta x>0$ i.e. there is a deficit.
The entropy change on the horizon due to the mass $m$ is then
$$\Delta S={{kc^3} \over 4G \hbar} \Delta A \simeq -{{2 \pi kmc} \over \hbar} \Delta x \eqno(8)$$
which is eq. (2), i.e. the formula postulated by Verlinde for the change in the entropy of a screen due to the motion of a mass close to it. Thus, Rindler horizons have all the properties of the screens postulated by Verlinde in ref. [\VER].{\footnote4{This fact was already noticed in ref. [\SUS].}}

This derivation clarifies the way a mass affects the entropy on the screen which in our case is the Rindler 
horizon. When the mass moves towards the horizon, i.e for $\Delta x<0$, the deflection angle remains the same but due to the geometry, the impact parameter $b^{\prime}$ on the horizon increases. Consequently, the area deficit decreases and the horizon entropy increases. When the mass moves away from the horizon the opposite occurs. This explains how the mass affects the horizon entropy without really interacting with it.
Note that, in the approximation we work, i.e. keeping the lowest order change in the area, the impact parameter disappears from the result in eq. (7). Therefore it can be as large as we like. Using the focusing of light, we calculated the finite difference between two infinite areas (of the Rindler horizon) that correspond to the mass at two different locations transverse to the horizon.
This method seems like an implicit way of regularizing the infinite horizon area and computing the finite area deficit that determines the entropy change.

We derived the Verlinde entropy formula in the weak gravity limit $Gm/b<<1$. In addition, we considered only the lowest order change to the area which means that we implicitly assumed $\Delta x << c^2b^2/2Gm$. This can be seen as a bound on $\Delta x$ for fixed impact parameter $b$. However, the assumption of weak gravity, i.e. $2Gm/bc^2<<1$ means the bound loosely becomes $\Delta x \leq b$. 
Since the impact parameter can be quite large, we find that $\Delta x$ does not necessarily have to be small, i.e. the mass does not necessarily have to be a Compton wavelength away from the screen as in Verlinde's treatment.

Another way to obtain the same result is to calculate the gravitational effect of the mass on the area of the horizon. The
Rindler horizon is the flat $y$--$z$ plane in the absence of the mass. However, in the presence of the mass space-time is described by the Schwarzschild metric and the horizon slightly deforms. An identical calculation that leads to the Verlinde entropy formula has been performed in ref. [\FUR].
There, a minimal surface which is a flat two-dimensional screen was placed close to a mass and the bending of the screen due to the gravitational field of the mass was computed. For weak gravity the Schwarzschild metric of the mass becomes
$$ds^2=- \left(1-{{2Gm} \over R} \right) dt^2+\left(1+{{2Gm} \over R} \right) (dx^2+dy^2+dz^2) \eqno(9)$$
The mass is at $y=z=0$ and $x=x_0$ and the horizon is the $y$--$z$ plane at $x=0$. 
Here, $R=\sqrt{x_0^2+r^2}$ (with $r^2=y^2+z^2$ the radius on the horizon) is the distance between an area element on the horizon and the mass. From eq. (9), we obtain the induced metric on the horizon (the $y$--$z$ plane) which leads to the horizon area in the presence of the mass 
$$A=\int r dr d \theta \left(1+{{2Gm} \over {(x_0^2+r^2)^{1/2}}} \right) \eqno(10)$$
From eq. (10) we find that if move the mass from $x=x_0$ by $\Delta x$ the area of the horizon changes by an amount
$$\Delta A \simeq - \Delta x \int r dr d\theta \left(1+{{2Gmx_0} \over {(x_0^2+r^2)^{3/2}}} \right)=-4 \pi Gm \Delta x \eqno(11)$$
which leads to half the desired result in eq. (7). In ref. [\FUR] this was rectified by taking two minimal surfaces (planes) on each side of the mass. Clearly this cannot be done in our case since there is only one Rindler horizon. The origin of this
discrepancy is not clear to us. In addition,
the result for area the depletion in eq. (11) is only the first term in an expansion in powers of $\Delta x$ whereas eq. (7) derived from the focusing of light has only a quadratic correction (which  was neglected in eq. (7)). Finally, the result in eq. (7) is independent of the impact parameter $b$ whereas eq. (10) requires an integral over the
whole horizon area, i.e. $r< \infty$. The bulk of the contribution to the area deficit comes from $r \leq \Delta x$ but there
is still a nonnegligible contribution from larger values. We see that the two methods of calculating the area depletion agree
(up to the factor of $1/2$) only at the lowest order.
It would be interesting to understand the relation between these two different ways of calculating the area depletion, find the origin of the missing factor of $1/2$ and resolve the discrepancy in the higher order contributions.

Once we have Verlinde's entropy formula, we can obtain Newton's second law following ref. [\VER]. Consider  lowering the mass
adiabatically towards the horizon. In this case, the external force acts away from the horizon (against gravitational pull of the horizon) and does work. On the other hand, this motion of the mass generates a change in the horizon entropy given by eq. (2). The mechanical work done to move the mass adiabatically is equal to the change in the thermodynamic energy of the Rindler horizon
$$F \Delta x=T \Delta S= {{\hbar a} \over {2\pi kc}} {{2 \pi k mc} \over \hbar} \Delta x=ma \Delta x \eqno(12) $$
where $T=\hbar a/ 2 \pi k c$ is the Rindler (or Unruh) temperature[\UNR] due to the acceleration $a$ and we used eq. (2) for $\Delta S$. Thus, we get $F=ma$ showing the entropic origin of Newton's second law.

The careful reader will probably notice that the work done by the external force, when lowering the mass towards the horizon, is negative while the change in horizon entropy and therefore the thermodynamic energy change is positive. Thus it seems that we
a sign discrepancy. However, the original direction of the acceleration in Minkowski space is opposite to the one that a mass at rest feels in Rindler space so in terms of the original acceleration Newton's second law holds as required.
We stress that the external force which we found to be equal to $ma$ is exactly the original force that accelerated the mass and is not necessarily gravitational.

Before we argue that gravity in Rindler space is an entropic force we need to clarify two issues. First,
since we used gravity, i.e. General Relativity, to compute the focusing of light in our derivation of the Verlinde formula, we cannot use eq. (7) to derive the gravitational force. Our results should be seen as a nontrivial consistency check between gravity in Rindler space, its entropic origin and the Verlinde entropy formula. In other words, the entropic origin of gravity is more manifest in Rindler space than in other backgrounds.

The second issue is the distinction between the gravitational field equations, i.e. how mass determines the gravitational field, and the equations of motion, i.e. how masses move in a gravitational field. The former, which can be described by Gauss' law or Einstein's equations, can be obtained solely from the thermodynamics of the horizon. In fact, Einstein's equations have been obtained from horizon thermodynamics some time ago[\JAC]. Below we
show the same for Newtonian gravity in Rindler space which is relevant for our purposes. On the other hand, the equations of motion, either $F=ma$ or the geodesic equation, can only be derived by using the entropic force in eq. (7). The content of the force is determined by the expression for the acceleration.

Now consider the Rindler horizon with entropy density per unit transverse area and Unruh or Rindler temperature given by
$${S \over A_{\perp}}={{kc^3} \over {4 \hbar G}} \qquad \quad T={{\hbar a} \over {2\pi kc}} \eqno(13)$$
Using the thermodynamics of the horizon, $E=2~TS$ we find the horizon energy density{\footnote5{The factor of 2 arises from the fact that for the energy one should use the Komar mass which includes pressure as a source of gravity[\PAD].}} 
$${E \over A_{\perp}}={{a c^2} \over {4 \pi G}} \eqno(14)$$
Thus, we can write for the gravitational field $a$
$$a A_{\perp}= 4\pi G \left(E \over c^2 \right)  \eqno(15)$$
This is Gauss' law for gravity in Rindler space (with constant acceleration and an infinite mass density on the horizon) which is
equivalent to the field equation in Newtonian gravity.
We stress that we have used, in addition to the Unruh temperature and holography, only the horizon thermodynamics. It is only when we want to write the gravitational force using the second law, $F=ma$, that we need the Verlinde entropy formula. 

The gravitational force that the horizon exerts on the mass is equal and opposite to the external entropic force in eq. (12), i.e.
$\vec F_{grav}=- \vec F$. This is why, in Rindler space, the mass is in equilibrium at a distance of $c^2/a$ from the horizon. Thus, using the same arguments as for $\vec F$, we can immediately conclude that the 
gravitational force is also given by eq. (12) and has an entropic origin. Alternatively, consider a mass freely
falling to the Rindler horizon due to gravity. This corresponds to a mass which is at rest in the original Minkowski space.
The work that gravity does when the mass falls by $\Delta x$ towards the horizon is equal to the change in the thermodynamic horizon energy as in eq. (12) (now without the sign discrepancy). In fact, this is precisely the change in the gravitational potential energy of the mass in Rindler space. Conventionally, we attribute the motion of the mass towards the horizon to the gravitational force
or equivalently to the gradient of the the gravitational potential energy. In the new entropic language, there is no gravity; the mass falls towards the horizon because its motion in that direction increases the horizon entropy. The gradient of the potential energy is replaced by the (negative of the) gradient of the horizon entropy.

The situation is very similar to a polymer in a heat bath that gives rise to a prototypical example of an entropic force[\VER]. In this case, every configuration of the polymer has the same energy but the one that maximizes the entropy
is the coiled state. If one were to extend the polymer and leave it, it would curl back due to the entropic force
that tries to maximize entropy which macroscopically looks like an elastic force that restores the coiled state. Similar comments can  be made for gravity in Rindler space.
The equilibrium state is the one in which the mass is merged with the horizon (either because it crossed it or fell into it and thermalized. The difference is immaterial for our purposes.) This is the state of maximum entropy due to the Verlinde entropy
formula in eq. (2). If we pull the mass out of the horizon to a certain distance and leave 
it, it freely falls to the horizon, i.e. to its equilibrium state with the maximum entropy just like the polymer curls when it is left in an extended state. Therefore, what macroscopically and conventionally looks like gravity is an entropic force.

An analogous argument can be made for General Relativity, i.e. Einstein's equations can be derived from the thermodynamics
of the Rindler horizon. In fact, this has been done a long time ago in ref. [\JAC] which also used (local) Rindler horizons. Here, for
completeness, we briefly review the derivation of ref. [\JAC] applied to global Rindler spaces.

Consider the Rindler horizon in the accelerated frame with the Killing field $\chi^{\mu}$. The energy or heat crossing the horizon due to infalling matter is given by (in the discussion below we set $c=k=1$)
$$\delta Q= \int_H T_{\mu \nu} \chi^{\mu} d\Sigma^{\nu} \eqno(16)$$
If $k^{\mu}$ is the tangent vector to the horizon generators and $\lambda$ is an affine parameter that vanishes on the horizon, then
$\chi^{\mu}=-a \lambda k^{\mu}$ and $d\Sigma^{\nu}=k^{\nu} d\lambda dA$ where $a$ and $A$ are the acceleration and area respectively. This gives
$$\delta Q=-a \int_H \lambda T_{\mu \nu} k^{\mu} k^{\nu} d\lambda dA \eqno(17)$$
Now, we assume holography on the Rindler horizon, i.e. $dS=dA/(4G \hbar)$. The variation of the area can be written as
$$\delta A= \int_H \theta d\lambda dA \eqno(18)$$
where $\theta$ is the expansion of the horizon generators whose change is determined by the Raychaudhuri equation
$${{d \theta} \over {d\lambda}}=-{1 \over 2} \theta^2-\sigma^2 -R_{\mu \nu} k^{\mu} k^{\nu} \eqno(19)$$
where $\sigma_{\mu \nu}$ and $R_{\mu \nu}$ are the shear and the Ricci tensors respectively. Now, $\theta,\sigma \sim R$ and for a small amount of energy crossing the horizon, we can neglect the first two terms on the right hand side since they are quadratic in $R_{\mu \nu}$.
Thus, we get $\theta=-\lambda R_{\mu \nu} k^{\mu} k^{\nu}$ for small $\lambda$, i.e. close to the horizon. Then, eq. (18) becomes
$$\delta A=-\int_H \lambda R_{\mu \nu} k^{\mu} k^{\nu} d\lambda dA \eqno(20)$$
Using the thermodynamic relation $\delta Q=T dS$, the Rindler temperature $T=\hbar a/2 \pi$ and the local conservation
of energy and momentum we get Einstein's equations
$$R_{\mu \nu}-{1 \over 2}R g_{\mu \nu}+\lambda g_{\mu \nu}=8 \pi G T_{\mu \nu} \eqno(21)$$

There is a subtlety about the derivation above that concerns the area and affine parameter integrals in eq. (17)[\JR]. The area integral has to range on a regularized (finite) horizon area since the actual area of the horizon is infinite. In addition, the range of the affine
parameter $\lambda$ also needs to be finite since for any mass, light rays coming off the horizon in the perpendicular direction,
i.e. the horizon generators, meet somewhere behind the mass at a caustic. This is a singularity of the Raychudhuri equation and the
range of $\lambda$ should not include the caustic. 

The fact that Rindler horizons serve as screens that satisfy the Verlinde formula  was already noticed in refs. [\CUL,\LEE]. However, the derivation of the formula that appears in these works though technically correct is conceptually wrong. We can now clarify this
point and show how the formulas worked out even though the ideas were incorrect. In refs. [\CUL,\LEE], eq. (2) was derived by the following argument. Consider a particle of mass $m$ created from the horizon. Then, the loss of energy on the horizon must equal the thermodynamic energy change there. Thus $mc^2=T \Delta S$. Using the formula for the Rindler (or Unruh) temperature and $a=c^2/\Delta x$ we get eq. (2). This derivation is not satisfactory because if we move the mass from its location $\Delta S$ changes whereas $mc^2$ is constant. 
The energy and entropy change of the screen (horizon) is due to the location of the mass and not its rest energy which is determined by its mass. Conversely, according to this prescription, the location of the mass is fixed so the mass cannot move. Therefore, it is not clear what $\Delta x$ means in eq. (2). Nevertheless, surprisingly, the above operations lead to the correct formula.

This situation can be clarified by taking into account the gravitational potential energy of the mass due to the finite mass density on the Rindler horizon. The uniform horizon energy density given by eq. (14) causes a uniform acceleration and in turn a linearly rising gravitational potential
$$V=4 \pi G \left ({{E/c^2} \over A_{\perp}}\right) \Delta x \eqno(22)$$
which is nothing but the work done per unit mass by adiabatically lifting the mass a distance $\Delta x$ above the horizon, i.e. 
$F \Delta x$ where the external force is equal and opposite the gravitational force between the horizon and the mass. The potential energy of the mass
at $\Delta x=c^2/a$ is exactly $mc^2$. Thus, in refs. [\CUL,\LEE], it was the gravitational potential energy, i.e. the work done against gravity that was set equal to the thermodynamic work. This seems to be the correct interpretation of the formulas in refs. 
[\CUL,\LEE].
With this interpretation, we can raise or lower the mass and both the gravitational potential energy and the thermodynamic horizon energy will change in equal but opposite amounts. Therefore 
$$ma \Delta x=-T \Delta S={{\hbar a} \over {2\pi kc}} \Delta S \eqno(23)$$
Solving for $\Delta S$ we get eq. (2) where now $\Delta x$ is arbitrary as required.
It is not clear to us whether the equality of the rest energy to the gravitational potential energy at $\Delta x=c^2/a$ is just
a coincidence or has a deeper meaning.

\bigskip
\centerline{\bf 3. Corollaries}
\medskip

In this section, we describe three results that arise directly from the Verlinde entropy formula, eq. (2). The first corollary is
the application of the formula to more general space-times such as the near horizon regions of the Schwarzschild and de Sitter
space-times (and black branes). Gravity should behave as an entropic force in these cases since these near horizon geometries reduce to Rindler space.
In addition, we show that eq. (2) directly implies the Bekenstein entropy bound for weakly gravitating bodies which can be seen as a new derivation of the bound. Finally, we discuss the connection of the area depletion to the uncertainty principle. 

{\it 3.1. Entropic Gravity in Near Horizon Geometries}~: 
We showed that, in Rindler space, the entropy formula (2) holds and gravity is an entropic force. Above, Rindler space was obtained by going into an accelerated frame in Minkowski space. On the other hand, it is well-known that Rindler space is also the near horizon geometry of Schwarzschild and de Sitter space-times in addition to those of all black objects such as black branes. Therefore, we expect gravity to behave as an entropic force also in these cases. For example, consider
the metric of a Schwarzschild black hole of mass $M$
$$ds^2=- c^2 \left(1-{{2GM} \over {rc^2}} \right) dt^2 + \left(1-{{2GM} \over {rc^2}} \right)^{-1} dr^2+ r^2 d \Omega_2^2 \eqno(24)$$
In the near horizon region defined by $r=R_s+y$ where $y<<R_s$ and $R_s=2GM/c^2$ is the Schwarzschild radius, the metric becomes that
of Rindler space with
$$ds^2=-\rho^2 d \tau^2 +d \rho^2 +d \Omega_2^2 \eqno(25)$$
where $\tau=t/2 R_s$ is the dimensionless Rindler time and the proper distance to the horizon is given by $\rho=2\sqrt{R_s y}$. Thus, an observer at rest at a fixed proper distance 
$\rho$ from the horizon sees Rindler space around her. This is not surprising since she is an accelerated observer.
The black hole horizon at $r=2GM/c^2$ coincides with the Rindler horizon at $\rho=0$ and the Unruh temperature is the Hawking temperature of the black hole. Therefore, we can immediately conclude that the Verlinde entropy formula holds in the near horizon region of Schwarzschild black holes and gravity behaves as an entropic force there.

The same idea can be applied to de Sitter space with the metric
$$ds^2=- c^2 \left(1-{{H^2r^2} \over c^2} \right)+ \left(1-{{H^2r^2} \over c^2} \right)^{-1}dr^2+r^2 d \Omega_2^2 \eqno(26)$$
In this case, the near horizon region is defined by $r=(c/H)+y$ where $y<<c/H$ and $c/H$ is the radius of the de Sitter space. This becomes Rindler space with the metric in eq. (25) where the proper distance to the horizon is $\rho=\sqrt{2Hy/c}$. Again, the   
the cosmological horizon at $r=c/H$ coincides with the Rindler horizon at $\rho=0$ as do the Unruh and de Sitter temperatures.
As expected, gravity is an entropic force near the de Sitter horizon.

We see that gravity behaves as an entropic force in the near horizon regions of Schwarzschild black holes and de Sitter spaces and the respective horizons to serve as screens. In fact, this is the case for all nonextreme objects including black branes since they all have a Schwarzschild factor in their metrics which guarantees that their near horizon regions are Rindler space. Showing this is fairly straightforward and we leave it to the reader.


It is difficult to generalize the above results to regions far from the horizon or to space-times with no horizons. The technical reason is that far from the horizon the space-times defined by the metrics in eqs. (24) and (26) do not become 
Rindler spaces. Therefore, our results do not apply there. It is certainly true that a static observer far from the black hole horizon is accelerated and in her rest frame she should see a Rindler space around her. However, the Rindler horizon, in this case, does not coincide with the black hole horizon and it is not clear how the two are related. 
In ref. [\VER], Verlinde derived Newtonian gravity and General Relativity in Minkowski space by assuming that the number of degrees of freedom on an arbitrary spherical screen surrounding a mass is saturated by the holographic bound. This is hard to justify since
it should only be true on horizons and not on hypothetical surfaces such as screens. 

{\it 3.2. The Bekenstein Entropy Bound}~:
We now derive the Bekenstein entropy bound that holds for weakly gravitating systems[\BEK] from a particular limit of the
Verlinde entropy formula. Consider
an object of radius $R$ and mass $m$ that accelerates with $a$. In the rest frame of this body there is a
Rindler horizon at a distance of $\Delta x =c^2/a$ from its center. 

Usually $\Delta x>>R$ since the acceleration is not too large. It is clear that the strongest bound on entropy will arise from the smallest possible $\Delta x$. On the other hand, 
we should demand that the Rindler horizon remain outside the object so that it is well-defined, i.e. $R \leq \Delta x$. Thus the smallest value is $\Delta x=R$ where the object touches the horizon. Using the Verlinde entropy formula, this immediately implies that there is a bound on the entropy of the object
$$S_{ob\!j} \leq \Delta S= {{2 \pi kmc} \over \hbar} R \eqno(27)$$
Using $mc=E/c$, this is precisely the Bekenstein entropy bound.
The inequality arises fron the fact that the process in which the object falls into the horizon increases the overall entropy.
Therefore the increase in the horizon (screen) entropy is an upper bound on the entropy of the object.
This derivation of the Bekenstein entropy bound is very similar to that in ref. [\BOU] which used the Generalized Covariant Entropy Bound[\GCEB] which is technically very similar to the above derivation of eq. (27) using sheets of light.
{\footnote6{For another derivation of the Bekenstein bound from the Bekenstein-Hawking entropy see ref. [\LAST].}}

{\it 3.3. The Uncertainty Principle}~:
Finally, we relate the uncertainty principle to the minimum area depletion.
This is not so suprising since the Bekenstein entropy bound implicitly contains the uncertainty principle as follows. Let us write
eq. (27) in a slightly more suggestive way as
$$S \leq 2 \pi k \left({{mc^2} \over {\hbar c/R}} \right) \eqno(28)$$
In the parenthesis, the numerator is the total energy of the object whereas the denominator is the smallest possible energy of a mode that fits into the region. The total number of such modes is the ratio and each mode carries an entropy of $2 \pi k$.

We now derive the uncertainty principle using the smallest possible area depletion which is $4 \ell_P^2=4 G \hbar/c^3$. From eq. (7)
we find the minimum area depletion due to the smallest mass (or energy) that fits in a region of size $L$ right above the horizon 
(with $\Delta x=L/2$)
$$\Delta A_{min}= {{8 \pi Gm} \over c^2} \Delta x \geq 4 \ell_P^2 \eqno(29)$$
Using the definition of the Planck length and $E=mc^2=pc$ for massless modes we find the lower bound on the momentum 
$p_{min} \geq {\hbar / {\pi L}}$ which is the uncertainty principle (up to a factor of $\pi$ that clearly arises from the circular shape of the areas in question). We note that here, the minimum area is not a small square with side $\ell_P$ but an annulus with a radius (approximately) equal to the impact parameter $b$ and width $4 \ell_P^2/b$. 

Alternatively, we can use the Bekenstein bound on the entropy of an object of mass $m$ in eq. (27) to find the minimum number of degrees of freedom that describes it
$$n={\Delta S \over {2 \pi k}} \leq {{E \Delta x} \over {\hbar c}} \eqno(30)$$
where we used the fact that each degree of freedom contributes $2 \pi k$ to the entropy. Since $n \geq 1$, using $E/c=p$ for massless
degrees of freedom we find
$$p~ \Delta x \geq \hbar \eqno(31)$$
which again is the uncertainty principle. Since above the Bekenstein bound was obtained from the Verlinde formula, this shows the relationship between the latter and the uncertainty principle.

The fact that we can relate the Verlinde entropy formula in eq. (2) to two very fundamental bounds of Nature such as the Bekenstein entropy bound and the uncertainty principle is additional though circumstantial evidence that eq. (2) describes an important
property of gravity.

\bigskip
\centerline{\bf 4. Speculations}
\medskip

In this section we entertain some speculative ideas that are related to gravity and its entropic nature in Rindler space.
The first speculation is about the existence of upper and lower bounds on acceleration due to the size of an object and the Hubble radius respectively.

Consider an object of radius $R$ with mass $m$ and acceleration $a$. We can always go to the accelerated frame (the rest frame of the object). In this frame, space becomes Rindler space with a Rindler horizon a distance $\Delta x=c^2/a$ from the center of mass of the object. However, we already argued above that we should demand $R \leq \Delta x$ so that the object is well-defined. In the previous section, this demand
led to the Bekenstein entropy bound which gives us confidence that it is an important physical condition. This immediately implies
an upper bound on the acceleration of the object, $a \leq a_{max}=c^2/R$. Note that the bound depends on the size of the object and is not universal.
On the other hand, we can also demand that the Rindler horizon be inside the cosmological horizon, i.e. the Hubble radius. This leads to a lower bound on acceleration, $a \geq a_{min}=Hc$ where $H$ is the Hubble constant. This bound is universal but time dependent due to
the time dependence of $H$. It is intriguing that this lower bound on acceleration is about the centripetal acceleration of galaxies,
i.e. $v^2/r \sim Hc$. This is exactly the scale at which the MOND hypothesis[\MIL,\BEKE] leads to the modification of Newton's second law in order to explain galactic rotation curves. Therefore, it would be interesting to find out whether the second law may be modified when the Rindler horizon approaches the Hubble horizon. Unfortunately, the concept of a Rindler horizon for rotating objects is not clear to us. 

Finally we raise the possibility of a holographic decription of Rindler horizons. From eq. (14) we see that the horizon energy density
per unit area satisfies $(E/A_{\perp}) \sim a \sim 1/\Delta x$, i.e. horizon energy is inversely proportional to the distance from
the screen. This is reminiscent of the well-known situation in AdS/CFT correspondence[\MAL,\GKP,\WIT]. However, the Rindler horizon corresponds to the IR region of space unlike the AdS boundary which corresponds to the UV region. Another difference with the AdS/CFT correspondence is the fact that in Rindler space, the horizon should parametrize the information about the space behind it which is not accessible to observers in front of it. In the AdS case the boundary is the end of space and therefore it parametrizes the physics in the bulk accessible to all observers (which is the difference between a boundary and a horizon).

We do not know what the theory that lives on the horizon is. However, clearly it has to have a constant holographic density of states and an energy density proportional
to temperature. In addition, it has to be a thermodynamic theory almost by assumption. 
Horizons that reduce to Rindler spaces in the near horizon region have been described by long strings with rescaled tensions in the past[\LENN-\EDI]. These do not seem to have the required properties listed above.
On the other hand, if the horizon theory were a gas of some kind, the fact that energy density is proportional to temperature would imply that it is a gas in $0+1$ dimensions, i.e quantum mechanics.
Recently an attempt to describe the horizon in terms of a matrix theory was made in ref. [\SUS].

\bigskip
\centerline{\bf 4. Conclusions and Discussion}
\medskip

In this paper, we derived the Verlinde entropy formula in Rindler space from the focusing of a sheet of light due to a mass infront of the Rindler horizon and the resulting area deficit. We also obtained the same result (up to a factor of two) from the gravitational effect of the mass on the area of the horizon. 
Thus, Rindler horizons are the entropic screens postulated by Verlinde in ref. [\VER] and gravity is an entropic force in
Rindler space. This result should be seen not as a derivation of gravity from eq. (2) but rather as a nontrivial consistency check between gravity in Rindler space, its entropic nature and the Verlinde formula. Our results also apply to space-times that reduce to Rindler space such as the near horizon geometries of Schwarzschild and de Sitter space-times and those of black branes.


Even though we have shown that gravity behaves as an entropic force in Rindler space and near horizon geometries, we have not been
able to generalize this result to other space-times especially those without horizons. As we argued above, it is difficult to describe gravity at a generic point
in space by a straightforward application of our methods due to the fact that the Rindler horizon that arises in the locally accelerating frame does not coincide with (and in fact is very far from) the point we want to describe.  
Needless to say, it is very important to generalize our results
to generic space-times with or without horizons and obtain Newtonian gravity and General Relativity. 

In ref. [\VER], Verlinde derived Newtonian gravity by assuming that spherical screens, i.e. Gaussian surfaces around masses, saturate the holographic entropy bound. This assumption is very hard to justify since it is true only on bona fide horizons which the screens,
in general, are not. This assumption seems to be implying that the screens behave as virtual black holes with much less mass than would be required from to their Schwarzschild radius. It is not clear why this should be the case. 

The situation may be better with respect to General Relativity since at any given point in curved space-time, we can go to the locally Rindler frame which describes an observer at rest at that point. Thus, it seems that our results about
Rindler space can be easily generalized to curved space-times. This is basically the content of refs. [\JAC,\PADD] which derived Einstein's equations from the thermodynamics on Rindler horizons. Clearly, this does not depend on the entropic nature of gravity but only on the horizon thermodynamics. As we discussed above, the field equations are a separate issue from the equations of motion. It would be interesting to see if the geodesic equation in curved space-times can be
obtained from local Rindler space arguments similar to the ones used in this paper.

We stress that, with respect to General Relativity described as an entropic effect, there are deep conceptual issues. 
For example, in General Relativity, gravity is not a force but simply a result of the curvature of space-time.
Even if we can obtain Einstein's equations from horizon thermodynamics, it is not clear what the connection between the entropy of screens and the curvature of space-time is. Naively, one would think that the geodesic direction would be the direction of
the steepest entropy change. In ref. [\VER] Verlinde derived the geodesic equations but again, this derivation suffers from the
hard to justify assumption we discussed above.

A more pessimistic possibility is that our results cannot be easily generalized to generic space-times. This would mean that
even though gravity is an entropic force, this property is only manifest in Rindler space. In fact, this would be very similar to the situation with respect to holography which is believed to be one of the fundamental properties of quantum gravity. Gravity may be holographic
in all space-times but this is, at least at the moment, manifest only in anti-de Sitter spaces through the AdS/CFT correspondence. It is interesting that anti-de Sitter spaces arise in the near horizon limit of extremal black holes whereas Rindler spaces arise in the near horizon limit of nonextremal black holes. Perhaps this connection may lead to a fusion of the concepts of holography and
entropic force.

\bigskip
\centerline{\bf Acknowledgments}

I would like to thank Lenny Susskind for very useful discussions and the Stanford Institute for Theoretical Physics for hospitality.

\vfill

\refout

\end
\bye